\documentclass[twocolumn,showpacs,preprintnumbers,amsmath,amssymb,prl,floatfix]{revtex4}
\usepackage{graphicx}% Include figure files
\usepackage{dcolumn}% Align table columns on decimal point
\usepackage{bm}% bold math
\usepackage{color}
\begin{document}

\title{
Flow-Induced Charge Modulation in Superfluid Atomic Fermions
Loaded into an Optical Kagome Lattice
}
%Flow-Induced Spatial Density Modulation of Superfluid Fermi Gases in a optical kagome Lattice

\author{Daisuke Yamamoto$^{1}$}
\author{Chika Sato$^{2}$}
\author{Tetsuro Nikuni$^{2}$}
\author{Shunji Tsuchiya$^{2}$}
\affiliation{
{$^1$Condensed Matter Theory Laboratory, RIKEN, 2-1 Hirosawa, Wako, Saitama 351-0198, Japan}
\\
{$^2$Department of Physics, Faculty of Science, Tokyo University of Science, 1-3 Kagurazaka, Shinjuku, Tokyo 162-8601, Japan}
}
\date{\today}% It is always \today, today,
             % but any date may be explicitly specified
\begin{abstract}
 We study the superfluid state of atomic fermions in a tunable optical
 kagome lattice motivated by recent experiments. We show that imposed
 superflow induces spatial modulations in the density and order parameter of
 the pair condensate and leads to a charge modulated superfluid state
 analogous to a supersolid state.  
 The spatial modulations in the superfluid emerge due to the geometric
 effect of the kagome lattice that introduces anisotropy in hopping amplitudes of fermion pairs in the presence of superflow. We also study superflow instabilities and find that the critical current limited by the dynamical instability is quite enhanced due to the large density of states associated with the flatband. 
 The charge modulated superfluid state can sustain high temperatures
 close to the transition temperature that is also enhanced due to the
 flatband, and is therefore realizable in experiments. 

\end{abstract}
\maketitle

%%%%%%%%%%%%%%%%%%%%%%%%%%%%%%%%%%%%%%%%%%%%%%%%%%%%%%%%%%%%%%%%%%%%%%%%%%%%%%%
%%%%%%%%%%%%%%%%%%%%%%%%%%%%%%%%%%%%%%%%%%%%%%%%%%%%%%%%%%%%%%%%%%%%%%%%%%%%%%%
%%                                                                           %%
%% Section I: introduction                                                   %%
%%                                                                           %%
%%%%%%%%%%%%%%%%%%%%%%%%%%%%%%%%%%%%%%%%%%%%%%%%%%%%%%%%%%%%%%%%%%%%%%%%%%%%%%%
%%%%%%%%%%%%%%%%%%%%%%%%%%%%%%%%%%%%%%%%%%%%%%%%%%%%%%%%%%%%%%%%%%%%%%%%%%%%%%%

Geometric frustration is a central subject in modern condensed matter
physics. Various ordered and liquid phases as well as multiferroic
behavior can arise from geometric frustration.
The kagome net is a well-known example of lattice geometry that exhibits
a high degree of frustration. This lattice geometry is proposed to host various exotic phases
such as a quantum spin liquid, valence bond solid \cite{Elser,Marston,Sachdev,Balents}, and the fractional quantum
Hall state \cite{Tang}. The intriguing feature of the kagome lattice is the
nondispersing flatband arising from geometric frustration. It
enhances the interaction effect and leads to ferromagnetic order 
\cite{Tasaki} as well as the destruction of Bose-Einstein condensation and
the resulting supersolid state~\cite{Huber}. Superconductivity (supefluidity) on the kagome lattice is a recent topic of
great interest from the theoretical side, despite the fact that few corresponding systems are known
\cite{Kiesel}. The infinitely large density of states associated with the
flatband can strongly enhance superconductivity in the kagome
lattice \cite{Imada}. Geometric frustration also provides a highly nontrivial effect on electron correlations and thus can lead to the emergence of novel superconducting states \cite{Hiroi}. 

\par
Ultracold atoms trapped within optical lattices offer an ideal system
for exploring various exotic phases and studying quantum phase
transitions due to their remarkable controllability and cleanness
\cite{Bloch}. 
The stability of superflow and critical velocity are of particular
interest for Fermi and Bose superfluids in optical lattices since the
pioneering experiments by Ketterle $et$ $al$. \cite{Ketterle}.
Various dynamical instabilities arising from competing orders in optical
lattices and the possibility of a ``flowing supersolid'' state have been investigated 
\cite{burkov-08,ganesh-09,yunomae-09,danshita-10,tsuchiya-12}.
The experiments by Jo {\it et al}.~\cite{jo-12} realized a tunable
optical lattice in the kagome geometry by overlaying two triangular
lattices with commensurate wavelengths. 
By introducing fermionic isotopes $^6{\rm Li}$ or $^{40}{\rm
K}$ into optical kagome lattices, this system would provide a very important
platform for studying superconductivity and Cooper pairing in the kagome lattice.

\begin{figure}[b]
\includegraphics[scale=0.28]{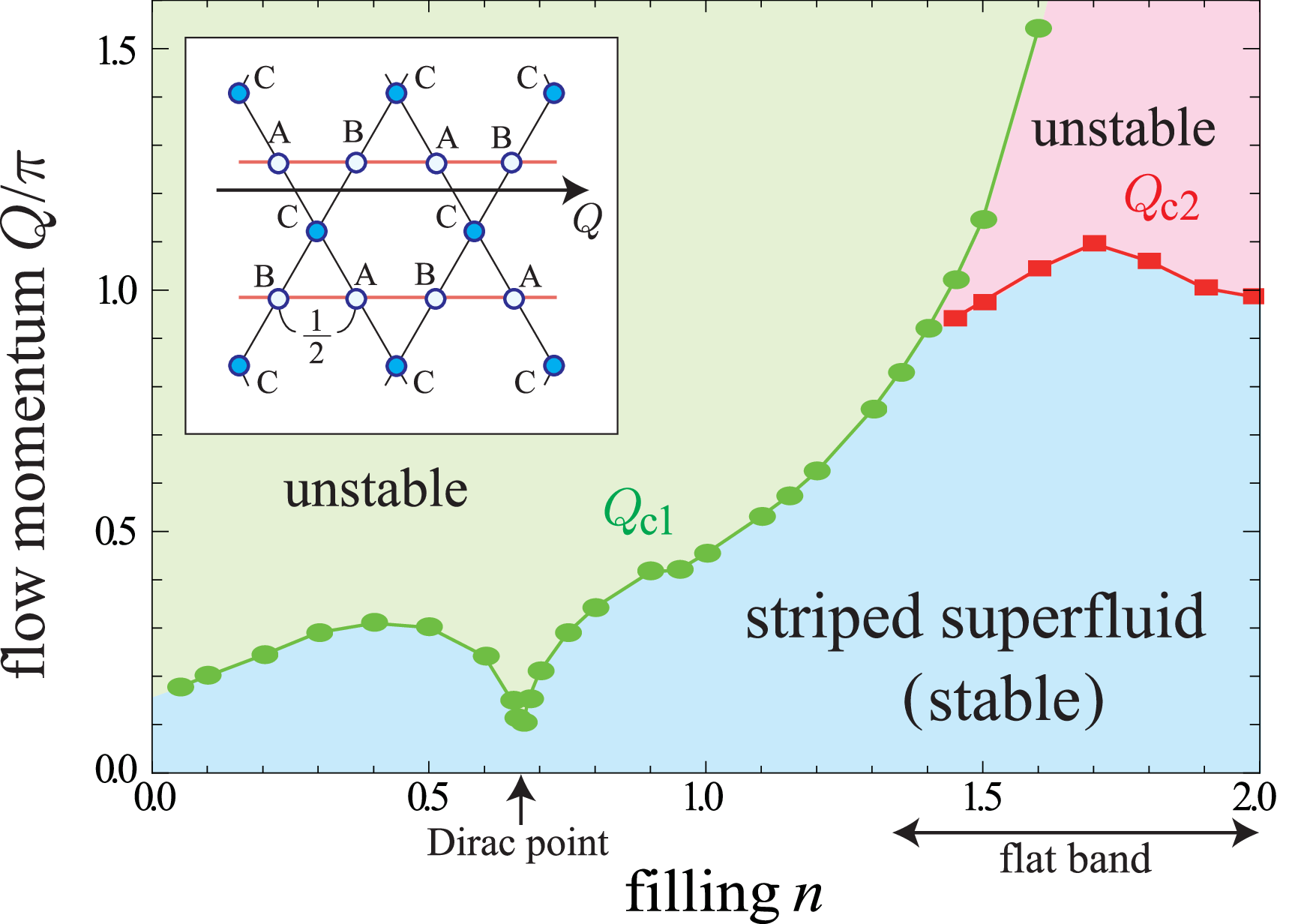}
\caption{\label{fig1} (color online).
Stability phase diagram of a flowing superfluid state in the kagome lattice.
We set the flow momentum ${\bf Q}=(Q,0)$ and $T=0$. 
$Q_{\rm c1}$ and $Q_{\rm c2}$ are the critical flow momenta for dynamical
instabilities with distinct mechanisms (see the text). 
The inset shows the stripe charge order induced by superflow with
$n_{ A}=n_{ B}\neq n_{ C}$ (see the text). 
} 
\end{figure}
\par
In this Letter, motivated by the recent experimental developments
reported by Jo {\it et al.}~\cite{jo-12}, we study the superfluidity of
atomic fermions in an optical kagome lattice within the attractive fermion Hubbard model.
Imposing a nonzero superflow, we calculate an order parameter
characterizing a condensate of fermion pairs
as well as fermion density within the mean-field approximation.
We further study superflow instabilities which limit the critical
current by evaluating the free energy as a function of the flow momentum. 
Figure~\ref{fig1} shows the stability phase diagram of an $s$-wave
superfluid state that summarizes the main results of this Letter.
We find a novel geometric effect of the kagome lattice that leads to a
stable flow-induced charge modulated superfluid state where the
superfluid order and charge density wave (CDW) order coexist (see the inset of
Fig.~\ref{fig1}) analogous to the ``flowing supersolid'' proposed in Ref.~\onlinecite{burkov-08}. 
Furthermore, the critical current for this state is found to be quite
enhanced at high fermion density due to the diverging density of states (DOS) associated with the flatband.
We also find that the instabilities at the critical current are dominated by
the different mechanisms at the low and high fermion densities. The
stability phase diagram thus shows a remarkable particle-hole
asymmetry of the critical current. 
The unexpected charge modulations we uncover can be
observed by employing the setup of the recent cold-atom
experiments~\cite{jo-12} combined with the moving optical lattice
technique~\cite{miller-07,fallani-04,Ketterle} or dipole
oscillations~\cite{burger-01fertig-05}. The charge modulations
induced by superflow may be also observed around a quantized vortex.
\par
%%%%%%%%%%%%%%%%%%%%%%%%%%%%%%%%%%%%%%%%%%%%%%%%%%%%%%%%%%%%%%%%%%%%%%%%%%%%%%%
%%%%%%%%%%%%%%%%%%%%%%%%%%%%%%%%%%%%%%%%%%%%%%%%%%%%%%%%%%%%%%%%%%%%%%%%%%%%%%%
%%                                                                           %%
%% Section II: model and method                                              %%
%%                                                                           %%
%%%%%%%%%%%%%%%%%%%%%%%%%%%%%%%%%%%%%%%%%%%%%%%%%%%%%%%%%%%%%%%%%%%%%%%%%%%%%%%
%%%%%%%%%%%%%%%%%%%%%%%%%%%%%%%%%%%%%%%%%%%%%%%%%%%%%%%%%%%%%%%%%%%%%%%%%%%%%%%
We consider two-component atomic fermions in a deep optical kagome lattice described by the attractive Hubbard model: $\hat{H}=
-t\sum_{\langle i,j \rangle,\sigma}
\left(\hat{c}^{\dagger}_{i\sigma} \hat{c}_{j\sigma}+{\rm h.c.}\right)
-U\sum_{ i }
 \hat{n}_{i\uparrow}\hat{n}_{i\downarrow}
-\mu \sum_{i,\sigma} \hat{n}_{i\sigma}$, 
where $\hat{c}^{\dagger}_{i\sigma}$ creates a fermion 
with spin $\sigma(=\uparrow,\downarrow)$ at site $i$,
$\hat{n}_{i\sigma}=\hat{c}^{\dagger}_{i\sigma} \hat{c}_{i\sigma}$, $t(>0)$ 
is a hopping amplitude between the nearest-neighbor sites, $U(\geq 0)$ is the
local Hubbard attraction, and $\mu$ is the chemical potential. For $U=0$, we obtain the three energy bands:
$\varepsilon_{0}=2t$ and $\varepsilon_{\pm}({\bf
k})=t\left(-1\pm\sqrt{3+2\sum_{i=1,2,3}\cos ({\bf k}\cdot {\bf
a}_i)}\right)$ with ${\bf a}_1=(1,0)$, ${\bf a}_2=(1/2,\sqrt{3}/2)$, and
${\bf a}_3={\bf a}_2-{\bf a}_1$. 
The flat band $\varepsilon_0$ is occupied if the fermion filling $n$ is in the range of
$4/3\leq n\leq 2$, while the Fermi energy is at the Dirac points in $\varepsilon_{\pm}({\bf
k})$ when $n=2/3$.
\par
We introduce the order parameter in the presence of superfluid
flow with flow momentum $\bf Q$ in the reference frame fixed to
the lattice potential \cite{EPAPS,frame}, 
$
\Delta_{\nu}=(U/M)\sum_{\rm k}  
\langle\hat{c}_{-{\bf k}+{\bf Q}/2,\nu,\downarrow}\hat{c}_{{\bf k}+{\bf
Q}/2,\nu,\uparrow}\rangle$,
and the average number of atoms per site,
$
n_\nu= (1/M)\sum_{{\bf
k},\sigma}\langle\hat{c}^\dagger_{{\bf k},\nu,\sigma}\hat{c}_{{\bf
k},\nu,\sigma}\rangle$,
where $\hat c_{\bf k,\nu,\sigma}$ is the Fourier component of $\hat
c_{i\sigma}$, $M$ is the number of unit cells, and $\nu$($=A,~B,~C$) labels the sublattice as shown in
the inset of Fig.~\ref{fig1}. 
In the following calculation, we use the periodic boundary
condition assuming that the system size is sufficiently large ($M\gg 1$) so that
the momentum can take continuous values. 
The fermion filling is given by $n=(n_{ A}+n_{ B}+n_{ C})/3$.
Within the Hartree-Fock-Gor'kov (HFG) approximation, the Hamiltonian takes the form
\begin{eqnarray}
\hat{H}_{\rm HFG}=
\sum_{\bf k}\Psi_{\bf Q}^\dagger ({\bf k})\hat{h}_{\bf Q} ({\bf k})\Psi_{\bf Q} ({\bf k})+E^0_{\bf Q},
\label{hamiltonianHFG}
\end{eqnarray}
with 
\begin{eqnarray*}
\Psi_{\bf Q}^\dagger ({\bf k})=(\hat{c}_{{\bf k}_+,{ A},\uparrow}^\dagger~\hat{c}_{{\bf k}_-,{ A},\downarrow}~\hat{c}_{{\bf k}_+,{ B},\uparrow}^\dagger~\hat{c}_{{\bf k}_-,{ B},\downarrow}~\hat{c}_{{\bf k}_+,{ C},\uparrow}^\dagger~\hat{c}_{{\bf k}_-,{ C},\downarrow})
\end{eqnarray*}
and $E^0_{\bf Q}=-3M(\mu+Un/2)+M\sum_\nu (|\Delta_{\nu}|^2/U+Un_\nu^2/4)$. Here, ${\bf k}_\pm$ denotes $\pm {\bf k}+{\bf
Q}/2$. The explicit form of $\hat{h}_{\bf Q} ({\bf k})$ is presented in
the Supplemental Material~\cite{EPAPS}. We obtain the excitation
spectrum of Bogoliubov quasiparticles $E_{{\bf Q},\tau}({\bf k})$ for
the energy band $\tau$($=\pm,0$) as the eigenvalues of the matrix $\hat{h}_{\bf Q} ({\bf k})$.
To evaluate $\Delta_{\nu}$ and $n_\nu$, we solve three gap and three number equations self-consistently in the standard manner \cite{tsuchiya-12}.
This scheme is known to interpolate the BCS and BEC regimes at low temperatures \cite{Leggett}.
\begin{figure}[t]
\includegraphics[scale=0.36]{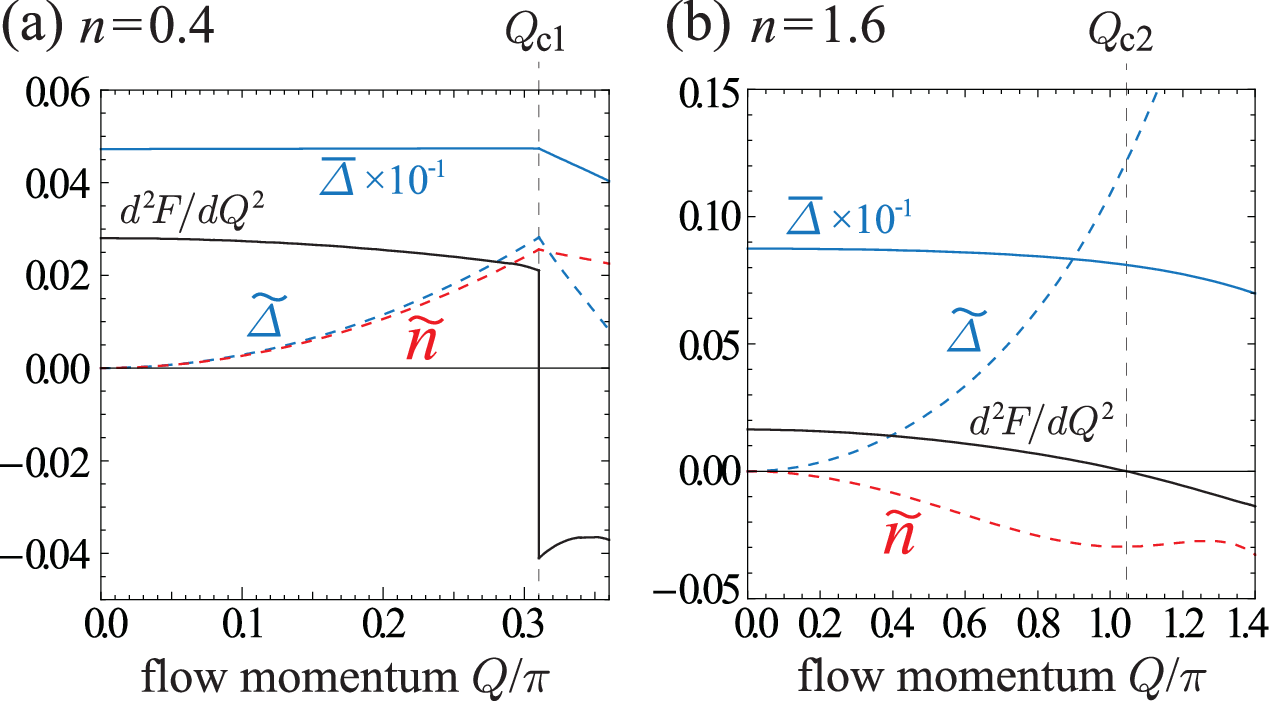}
\caption{\label{fig2} (color online). 
Density modulation $\tilde n$, order parameter
modulation $\tilde \Delta$, and averaged order parameter $\bar\Delta$ as
functions of flow momentum ${\bf Q}=(Q,0)$ for (a) $n=0.4$ and (b) $n=1.6$.
We set $U/t=3$ and $T=0$. $Q_{\rm c1}$ and $Q_{\rm c2}$ are the critical flow momenta. 
} 
\end{figure}
\par
In the stationary ground state ($\bf Q=0$), the order parameters as well
as the filling factors take the same value for all the sublattices by the symmetry.
If we impose nonzero superflow, we find that an infinitesimally small amount of superflow breaks this
symmetry and leads to spatial modulations in density and order parameter. 
We plot $\tilde{n}\equiv n_{ C}- n_{ A}$, $\tilde{\Delta}\equiv \Delta_{C}-\Delta_{A}$, and the averaged order parameter 
$\bar{\Delta}\equiv (\Delta_{A}+\Delta_{ B}+\Delta_{ C})/3$ 
as functions of $Q$ for different fillings in Figs.~\ref{fig2}(a) and~\ref{fig2}(b) setting the flow in the $\Gamma- K$ direction (${\bm Q}=(Q,0)$). 
The system involves a stripe order in the direction perpendicular to the flow with $\Delta_{ A}=\Delta_{ B}\neq \Delta_{C}$ and $n_{A}=n_{ B}\neq n_{ C}$ (see inset of Fig.~\ref{fig1}). 
This superfluid state with a stripe charge order is analogous to a supersolid state in
the sense that superfluid and CDW orders coexist.
Such a flow-induced charge modulated state arises due to the geometric effect that is
unique to the kagome lattice, while the supersolid state arises due to the spontaneous breaking of translational symmetry.

To check the stability of the superfluid state with stripe charge order,
we evaluate the quantity $\frac{1}{N}\frac{d^2 F(Q)}{dQ^2}$ that
represents phase stiffness~\cite{burkov-08}, where $N=3M$ is the number of total lattice sites and $F(Q)$ is
the free energy of the system in the presence of superflow.
If this value is negative, the system is dynamically unstable against
phase and density fluctuations~\cite{ganesh-09,burkov-08} 
so that the critical momentum of superflow is given by the value of $Q$ at
which $\frac{d F(Q)}{dQ}$ takes the
maximum~\cite{Goren}.
Note that the superfluid density is proportional to $d^2F(Q)/dQ^2|_{Q=0}$.
In Figs.~\ref{fig2}(a) and \ref{fig2}(b), $\frac{1}{N}\frac{d^2 F(Q)}{dQ^2}$ is positive for both the low and high fillings until the critical flow momentum $Q_{\rm c1}$ or $Q_{\rm c2}$. Consequently, the superfluid state with
the stripe charge order is dynamically stable until $Q_{\rm c1}$ or $Q_{\rm
c2}$. This is in sharp contrast with the ``flowing supersolid'' state in a
square lattice~\cite{ganesh-09}, which is found to be dynamically
unstable with negative $\frac{1}{N}\frac{d^2 F(Q)}{dQ^2}$ for any nonzero $Q$.
\begin{figure}[t]
\includegraphics[scale=0.3]{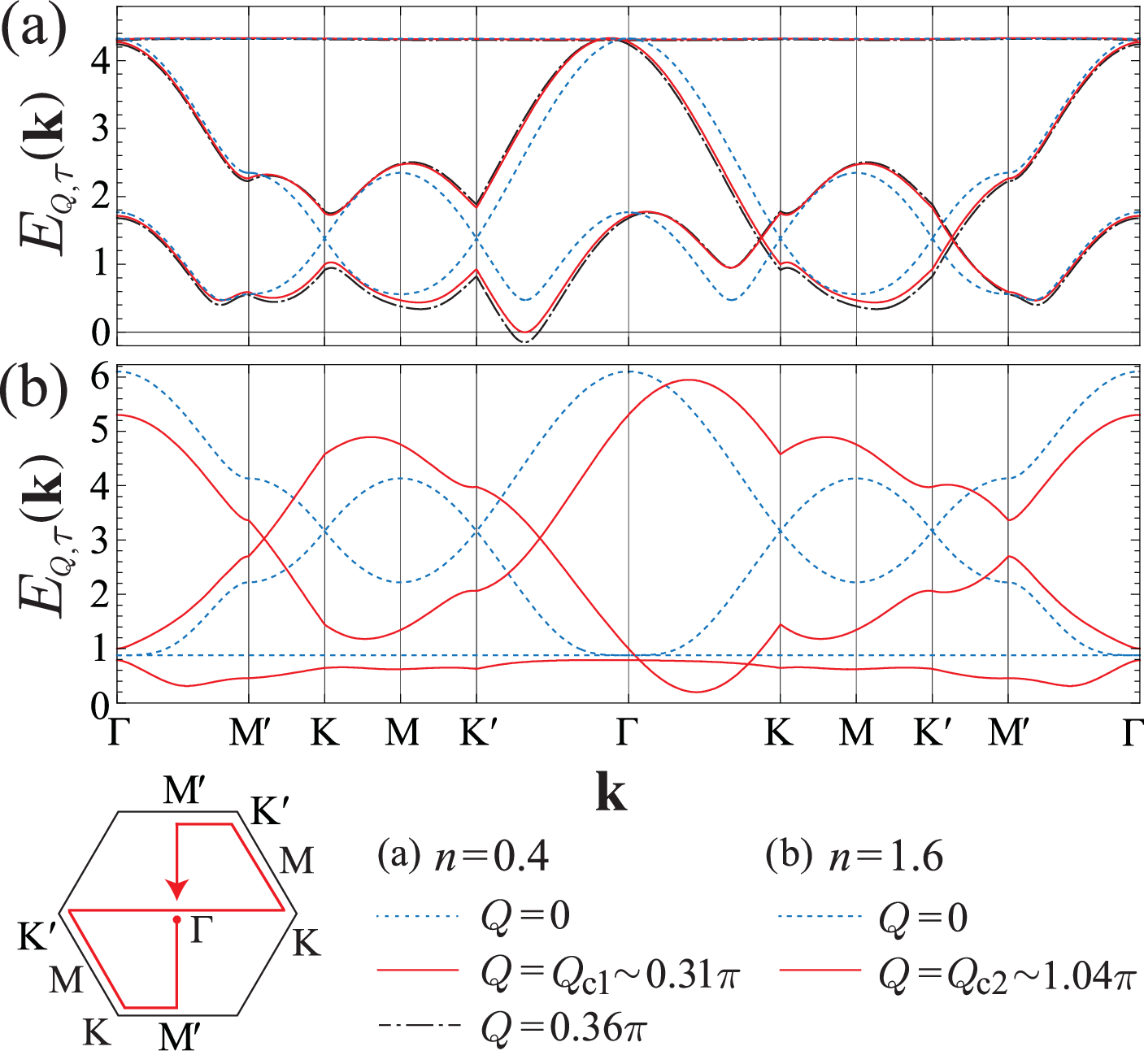}
\caption{\label{fig3} (color online). 
Quasiparticle energy band $E_{{\bf Q},\tau}({\bf k})$ ($\tau=0,\pm$) for
different values of imposed flow $\bm Q=(Q,0)$ at (a) $n=0.4$ and (b) $n=1.6$. 
We set $U/t=3$ and $T=0$. 
} 
\end{figure}
\par
We find that the dynamical instabilities are caused by different mechanisms for low and high fillings. 
The value of $\frac{1}{N}\frac{d^2 F(Q)}{dQ^2}$ in Fig.~\ref{fig2}(a) changes discontinuously
at the critical momentum $Q_{\rm c1}$, while the curve in Fig.~\ref{fig2}(b) smoothly 
changes from positive to negative at $Q_{\rm c2}$. 
We find that the sudden change of the curve at $Q_{\rm c1}$ is associated with
gapless quasiparticle excitations. With increasing flow, the energy gap to creating quasiparticles decreases due to a Doppler shift in the direction
opposite to $\bf Q$ as shown in Fig.~\ref{fig3} (a). 
For large $Q$, the gap closes and the lowest quasiparticle energy
band $E_{{\bf Q},\tau}({\bf k})$ becomes gapless. 
This precisely coincides with $Q_{\rm c1}$. 
Since the free energy at $T=0$ involves the contribution from
spontaneously excited quasiparticles with negative energies above the
critical flow $Q\geq Q_{\rm c1}$, the second derivative of the free energy $\frac{1}{N}\frac{d^2 F(Q)}{dQ^2}$ changes discontinuously at $Q=Q_{\rm c1}$.
On the other hand, in Fig.~\ref{fig2}(b), the quasiparticle
dispersion is still gapped at $Q_{\rm c2}$. Thus, the instability at
$Q_{\rm c2}$ sets in before the closing of the single-particle
excitation gap. The negative value of $\frac{1}{N}\frac{d^2 F(Q)}{dQ^2}$ therefore
indicates the dynamical instability associated
with collective phonon excitations rather than single-particle
excitations~\cite{burkov-08,machholm-03,taylor-03,pitaevskii-05}.
At the onset of the instabilities at $Q_{\rm c1}$ and $Q_{\rm c2}$, the frequency of
long-wavelength phonons becomes complex. As a result, the amplitude of
collective phonon excitations grows exponentially and the superfluid state collapses.
\par
The stability phase diagram in Fig.~\ref{fig1} exhibits a remarkable particle-hole asymmetry reflecting the
different features of $\frac{1}{N}\frac{d^2 F(Q)}{dQ^2}$ at $Q_{\rm c1}$ and $Q_{\rm c2}$
discussed above.
The critical momentum at high filling ($n \gtrsim 4/3$) is significantly
enhanced while being limited by the onset of the dynamical instability at
$Q_{\rm c2}$. The robust superfluidity against the imposed superflow is due to the flatband in the noninteracting
band structure. The diverging DOS at the flatband enhances the order parameter
and the single-particle energy gap. The large energy gap suppresses the
depairing instability since a large Doppler shift is required
for closing the gap. 
On the other hand, the small order parameter for low filling yields the
small energy gap in Fig.~\ref{fig3}. The dynamical instability at
$Q_{\rm c1}$ for low filling ($n\lesssim 4/3$) is therefore preempted by the closing of the
single-particle excitation gap.
\begin{figure}[t]
\includegraphics[scale=0.4]{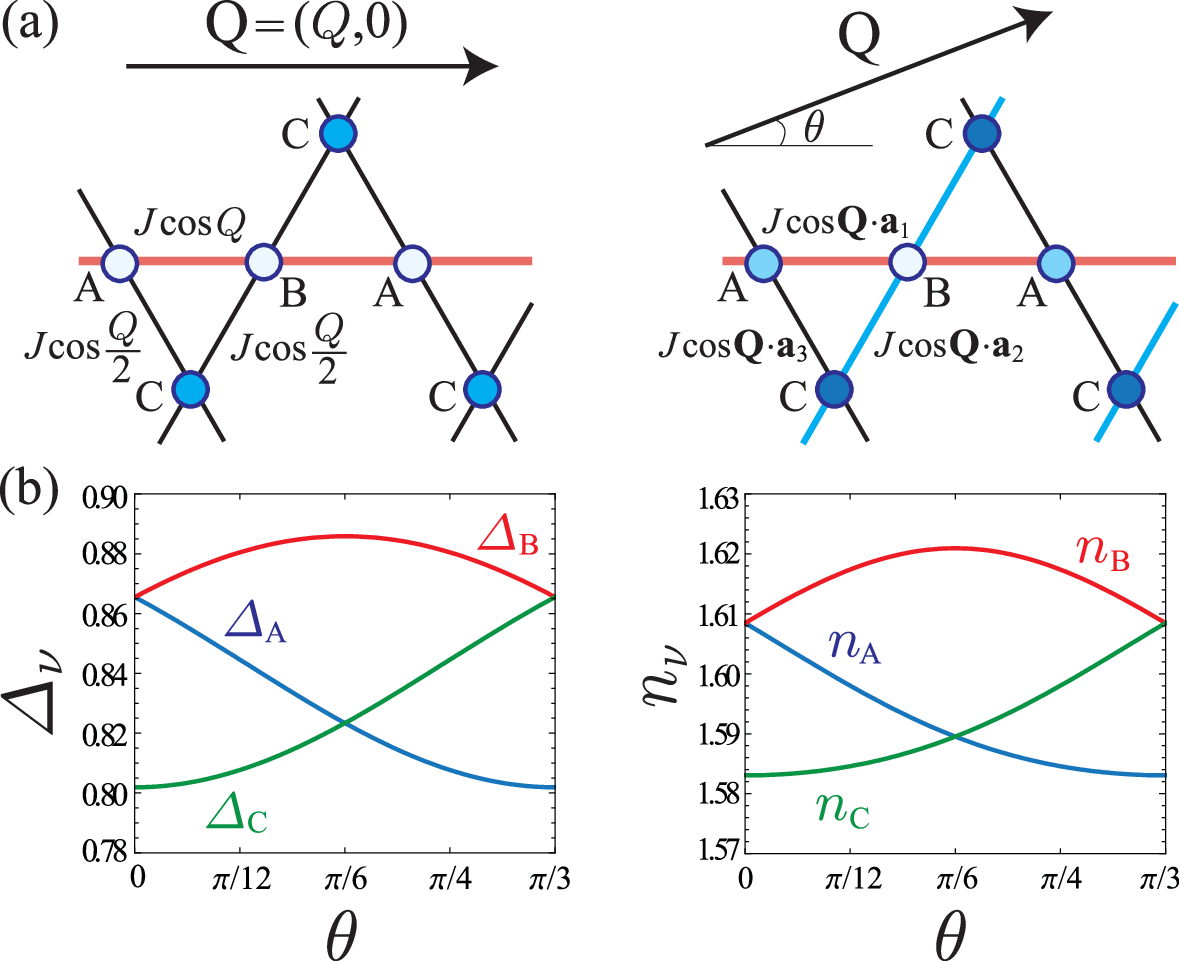}
\caption{\label{fig4} (color online). 
(a) Effective hopping amplitudes for fermion pairs in the strong coupling
regime when ${\bf Q}=(Q,0)$ and ${\bm Q}=(Q\cos\theta,Q\sin\theta)$. 
(b) Order parameter $\Delta_{\nu}$ and fermion filling
$n_{\nu}$ as functions of the angle $\theta$ defined in (b). We set $Q=0.8\pi$,
$n=1.6$, and $U/t=3$.
} 
\end{figure}
\par
We now discuss the reason for the emergence of the flow-induced
charge modulations.
To make the argument simpler, it is convenient to describe the
system in the rest frame of the condensate where the lattice potential
is moving with the velocity ${\bf v}=-{\bf Q}/2m$ \cite{frame} and we restrict
ourselves within the strong coupling regime ($U\gg t$) where fermion pairs become tightly bound
molecular bosons. 
In this regime, the hopping term of the effective Hamiltonian for bosons
in the presence of superflow ${\bf Q}$ is given by
$-J \sum_{\langle i,j \rangle} (e^{-i{\bf Q}\cdot {\bf 
r}_{ij}}b_i^\dagger b_j+{\rm H.c.})$. 
Here, $J=2t^2/U$, $b_i=\hat{c}_{i\downarrow}\hat{c}_{i\uparrow}$ is an annihilation
operator for a boson, and ${\bf r}_{ij}={\bf r}_{i}-{\bf r}_{j}$ is the
bond vector.
The Hamiltonian shows that the imposed superflow reduces the effective
hopping amplitude of bosons by 
a factor of $\cos{\bf Q}\cdot{\bf r}_{ij}$, which plays a crucial role for the spatial modulation.
The kagome lattice has three kinds of nearest-neighbor bonds connecting
two sites: the $A$-$B$, $B$-$C$, and $C$-$A$ bonds. 
The flow breaks the symmetry of the three bonds and introduces the
anisotropy in the hopping amplitude, which naturally leads to the
spatial modulation in density and order parameter.
For example, when the flow is in the $\Gamma\rightarrow K$ direction, the effective hopping amplitudes are given by
$J_{ AB}=J\cos {Q}$ for the $A$-$B$ bond and $J_{ BC}=J_{
CA}=J\cos \frac{Q}{2}$ for the $B$-$C$ and $C$-$A$ bonds as shown in Fig.~\ref{fig4}(a). 
Because of this anisotropy, the system prefers forming a stripe modulation with
$n_{ A}=n_{ B}\neq n_{ C}$ and $\Delta_{ A}=\Delta_{
B}\neq \Delta_{ C}$ in order to maximize the energy gain.
Figure~\ref{fig4}(b) shows $\Delta_{\nu}$ and $n_{\nu}$ as
functions of the angle $\theta$ of the flow momentum ${\bf Q}$ relative
to the $x$ axis. In general, $\Delta_{\nu}$ and $n_{\nu}$ 
take different values depending on sublattices $A$, $B$, and $C$ due to the anisotropy of the effective hopping amplitudes of pairs, except for some special symmetric points,
$\theta=0,\pi/6,\pi/3,\cdots$, where two of them are equivalent and the
system forms a stripe pattern.   
This spatial modulation obviously emerges from the characteristic geometry of the
kagome lattice and therefore it is absent in other typical lattice geometries such as a square~\cite{yunomae-09,ganesh-09}, cubic~\cite{yunomae-09,ganesh-09}, honeycomb~\cite{tsuchiya-12}, and even triangular lattice that has geometric frustration.

\par
\begin{figure}[t]
\includegraphics[scale=0.28]{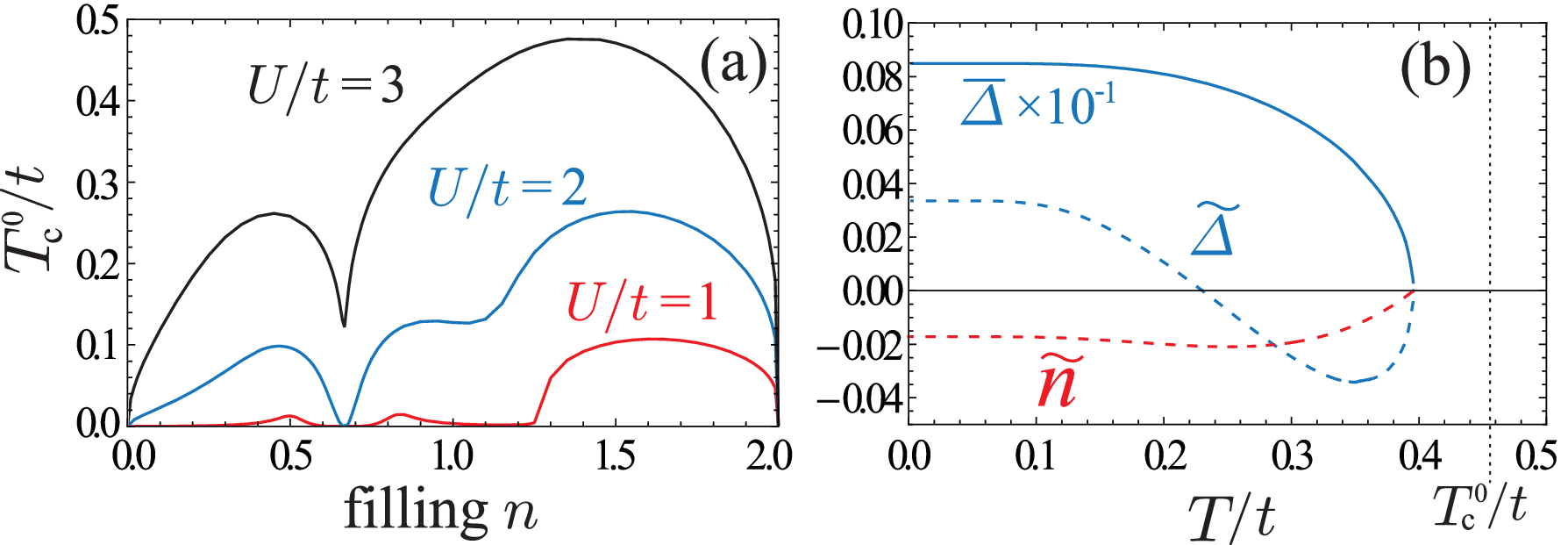}
\caption{\label{fig5} (color online) 
(a) Mean-field transition temperature $T^0_{\rm c}$ as a function of fermion
filling $n$. (b) Temperature dependence of $\tilde \Delta$, $\tilde n$, and $\bar \Delta$ for $n=1.6$, ${\bf Q}=(0.6\pi,0)$, and $U/t=3$. The vertical
 dotted line marks $T^0_{\rm c}$ in the absence of flow. 
} 
\end{figure}

For realizing a Fermi superfluid in optical lattices, 
experimental difficulties arise in cooling the system down to the superfluid
transition temperature. The kagome lattice has a great advantage in this respect.
The mean-field transition temperature $T^0_{\rm
c}$ in Fig.~\ref{fig5}(a) is significantly enhanced for high filling $n\gtrsim
4/3$ when the Fermi level reaches the flatband.
Despite the fact that a pure 2D system has no real
condensate of pairs at finite temperatures \cite{Mermin},
$T_c^0$ is useful for estimating the Kosterlitz-Thouless transition
temperature $T_{\rm KT}$ \cite{Kosterlitz} in the case of $T_F\gg T_c^0$, 
i.e., $(T_c^0-T_{\rm KT})/T_c^0\sim (T_c^0/T_F)\ll 1$ \cite{Miyake,Orso}.
$T_c^0$ also provides a good estimate for the actual transition
temperature $T_c$ of fermions in weakly coupled layers of kagome
lattices in which phase fluctuation that 
destroys condensate is suppressed due to the interlayer Josephson coupling while
the system maintains two-dimensional features.
Such a system can be realized by loading fermions into a series of
pancake-shaped potentials \cite{Martiyanov} together with kagome
optical lattices as realized in Ref.~\onlinecite{jo-12}.
Figure~\ref{fig5}(b) shows that the stripe charge order is observable up to high
temperatures slightly below $T^0_{\rm c}$. 
We note that the curve for $T^0_{\rm c}$ in Fig.~\ref{fig5}(a) shows a dip
in the vicinity of $n=2/3$ due to the Dirac points in the free
fermion band structure~\cite{zhao-06,lee-09,tsuchiya-12}. 
The system remains in the normal semimetallic phase
for small $U$ less than the critical value $U_{\rm c}\approx 2.8t$.
\par
The flow-induced charge modulations that we uncovered can be realized
in cold-atom experiments if superflow is imposed by a moving optical
lattice~\cite{miller-07,fallani-04,Ketterle}. 
The experiment in Ref.~\onlinecite{jo-12} overlays two triangular
lattices to form the kagome lattice. Each triangular lattice is formed
by three lasers at angles of 120 degrees with respect to each
other. Superflow can be induced in the $\Gamma-K$ direction by detuning
one of the lasers for each triangular lattice and moving both the
triangular lattices with the same velocity \cite{EPAPS}. Superflow can be also imposed by
dipole oscillations which can be induced by a sudden displacement of a
confining harmonic potential~\cite{burger-01fertig-05}.
Since infinitesimally small flow can induce charge
modulations, superflow around a single vortex induces charge modulations
that extend over the whole system; therefore, it may also be easily observed.

%%%%%%%%%%%%%%%%%%%%%%%%%%%%%%%%%%%%%%%%%%%%%%%%%%%%%%%%%%%%%%%%%%%%%%%%%%%%%%%
%%%%%%%%%%%%%%%%%%%%%%%%%%%%%%%%%%%%%%%%%%%%%%%%%%%%%%%%%%%%%%%%%%%%%%%%%%%%%%%
%%                                                                           %%
%% Section IV: finite superflow                                              %%
%%                                                                           %%
%%%%%%%%%%%%%%%%%%%%%%%%%%%%%%%%%%%%%%%%%%%%%%%%%%%%%%%%%%%%%%%%%%%%%%%%%%%%%%%
%%%%%%%%%%%%%%%%%%%%%%%%%%%%%%%%%%%%%%%%%%%%%%%%%%%%%%%%%%%%%%%%%%%%%%%%%%%%%%%

%%%%%%%%%%%%%%%%%%%%%%%%%%%%%%%%%%%%%%%%%%%%%%%%%%%%%%%%%%%%%%%%%%%%%%%%%%%%%%%
%%%%%%%%%%%%%%%%%%%%%%%%%%%%%%%%%%%%%%%%%%%%%%%%%%%%%%%%%%%%%%%%%%%%%%%%%%%%%%%
%%                                                                           %%
%% Section V: summary                                                        %%
%%                                                                           %%
%%%%%%%%%%%%%%%%%%%%%%%%%%%%%%%%%%%%%%%%%%%%%%%%%%%%%%%%%%%%%%%%%%%%%%%%%%%%%%%
%%%%%%%%%%%%%%%%%%%%%%%%%%%%%%%%%%%%%%%%%%%%%%%%%%%%%%%%%%%%%%%%%%%%%%%%%%%%%%%
In summary, we have studied the $s$-wave superfluid state of atomic fermions on the
kagome lattice inspired by the recent realization of
tunable kagome optical lattices~\cite{jo-12}. We performed a mean-field
analysis of the superfluid state imposing superflow. We found
that superflow induces a novel charge modulated state due to the
characteristic geometry of the kagome lattice. The superfluid and CDW
orders coexist in this state analogous to the ``supersolid'' state.  
We examined the superflow instabilities and critical current by
evaluating the free energy as a function of superflow.
The critical current for high filling was found to be quite enhanced
due to the flatband in the free fermion band structure.
The superfluid state with charge modulations sustains high
temperatures close to the mean-field transition temperature $T^0_{\rm
c}$ which is also enhanced by the flat band for high filling and
therefore accessible using the setup that is realizable in cold-atom experiments.
%%%%%%%%%%%%%%%%%%%%%%%%%%%%%%%%%%%%%%%%%%%%%%%%%%%%%%%%%%%%%%%%%%%%%%%%%%%%%%%
%%%%%%%%%%%%%%%%%%%%%%%%%%%%%%%%%%%%%%%%%%%%%%%%%%%%%%%%%%%%%%%%%%%%%%%%%%%%%%%
%%                                                                           %%
%% Section VI: acknowledgements                                              %%
%%                                                                           %%
%%%%%%%%%%%%%%%%%%%%%%%%%%%%%%%%%%%%%%%%%%%%%%%%%%%%%%%%%%%%%%%%%%%%%%%%%%%%%%%
%%%%%%%%%%%%%%%%%%%%%%%%%%%%%%%%%%%%%%%%%%%%%%%%%%%%%%%%%%%%%%%%%%%%%%%%%%%%%%%
\par
We acknowledge T. Ohkane for performing preliminary calculations in the
early stage of this work. S. T. and D. Y. were supported by a Grant-in-Aid for Scientific Research,
Grants No. 24740276 (S. T.) and No. 23840054 (D. Y.).

\widetext

\subsection{\large Supplementary Material for ``Flow-Induced Spatial Modulation of Superfluid Atomic Fermions in an Optical Kagome Lattice''}
\renewcommand{\thesection}{\Alph{section}}
\renewcommand{\thefigure}{S\arabic{figure}}
\renewcommand{\thetable}{S\Roman{table}}
\setcounter{figure}{0}
\subsection{\label{1s}A. The explicit form of $\hat{h}_{\bf Q} ({\bf k})$} 
The explicit form of the $6\times 6$ matrix $\hat{h}_{\bf Q} ({\bf k})$ in Eq.~(1) of the main text is given by
\begin{eqnarray*}
\hat{h}_{\bf Q}=\left(\begin{array}{cccccc}
		-\mu-\frac{U}{2}n_{\rm A}&-\Delta_{\rm A}&-2t\cos ({\bf k}_+\cdot \frac{{\bf a}_1}{2})&0&-2t\cos ({\bf k}_+\cdot \frac{{\bf a}_3}{2})&0\\
		-\Delta^\ast_{\rm A}&\mu+\frac{U}{2}n_{\rm A}&0&2t\cos ({\bf k}_-\cdot \frac{{\bf a}_1}{2})&0&2t\cos ({\bf k}_-\cdot \frac{{\bf a}_3}{2})\\
		-2t\cos ({\bf k}_+\cdot \frac{{\bf a}_1}{2})&0&-\mu-\frac{U}{2}n_{\rm B}&-\Delta_{\rm B}&-2t\cos ({\bf k}_+\cdot \frac{{\bf a}_2}{2})&0\\
		0&2t\cos ({\bf k}_-\cdot \frac{{\bf a}_1}{2})&-\Delta^\ast_{\rm B}&\mu+\frac{U}{2}n_{\rm B}&0&2t\cos ({\bf k}_-\cdot \frac{{\bf a}_2}{2})\\
		-2t\cos ({\bf k}_+\cdot \frac{{\bf a}_3}{2})&0&-2t\cos ({\bf k}_+\cdot \frac{{\bf a}_2}{2})&0&-\mu-\frac{U}{2}n_{\rm C}&-\Delta_{\rm C}\\
		0&2t\cos ({\bf k}_-\cdot \frac{{\bf a}_3}{2})&0&2t\cos ({\bf k}_-\cdot \frac{{\bf a}_2}{2})&-\Delta^\ast_{\rm C}&\mu+\frac{U}{2}n_{\rm C}
		\end{array}\right).
\end{eqnarray*}
The matrix $\hat{h}_{\bf Q}$ can be diagonalized by the Bogoliubov
transformation in the standard manner~\cite{tsuchiya-12s}. Without
imposed superflow, we can take $\Delta_{\rm
A}=\Delta_{\rm B}=\Delta_{\rm C}\equiv\Delta$ and $n_{\rm A}=n_{\rm
B}=n_{\rm C}\equiv n$ due to the symmetry of the lattice. In this case,
$\hat{h}_{\bf Q}$ can be analytically diagonalized to give the gap and
number equations
\begin{eqnarray}
\frac{\Delta}{U}=\frac{1}{N}\sum_{\bf k}\sum_{\tau =0,\pm}\frac{\Delta}{2E_{{\bf Q}={\bf 0},\tau}({\bf k})}\tanh\frac{\beta E_{{\bf Q}={\bf 0},\tau}({\bf k})}{2}\label{gap}
\end{eqnarray}
and
\begin{eqnarray}
n=1-\frac{1}{N}\sum_{\bf k}\sum_{\tau =0,\pm}\frac{\xi_{\tau}({\bf k})}{E_{{\bf Q}={\bf 0},\tau}({\bf k})}\tanh\frac{\beta E_{{\bf Q}={\bf 0},\tau}({\bf k})}{2},\label{number}
\end{eqnarray}
respectively. Here, $N=3M$ is the number of total lattice sites,
$\beta=1/T$ is the inverse temperature, and $\xi_{\tau}({\bf
k})=\varepsilon_{\tau}({\bf k})-\mu-Un/2$. The Bogoliubov quasiparticle
bands for ${\bf Q}={\bf 0}$ are simply given by $E_{{\bf Q}={\bf
0},\tau}({\bf k})=\sqrt{\xi_{\tau}({\bf k})^2+|\Delta|^2}$. 
There exists one flat band ($\tau=0$) even in the superfluid state
(see Fig.~3 of the main text).

\subsection{\label{2s}B. Moving kagome optical lattice} 
We present here how to prepare a {\it moving} optical lattice with
kagome geometry. In the recent experiment by Jo {\it et al}.~\cite{jo-12s}, the kagome
lattice was formed by overlaying two triangular optical lattices with
different lattice constants. Therefore, we only have to
consider the setup for moving a triangular-lattice potential with a
constant velocity. The triangular lattice is generated by superposing
three laser beams that intersect in the $x$-$y$ plane with wave vectors
${\bf k}_1$ = $k (1, 0)$, ${\bf k}_2$ = $k (-1/2, -\sqrt{3}/2)$, and
${\bf k}_3$ = $k (-1/2, \sqrt{3}/2)$. All beams are linearly polarized
orthogonal to the plane and have the same field strength $E_0$. The
total electric field is given by 
\begin{equation}
{\bf E}({\bf r}, t)=\sum_{i=1}^3E_0 \cos ({\bf k}_i\cdot {\bf r}-\omega t+\phi_i){\bf e}_z. 
\end{equation}
The relative phases $\phi_{ij}=\phi_{i}-\phi_{j}$ are fixed in Ref.~\onlinecite{jo-12s} to obtain a stable optical lattice. The generated dipole potential is proportional to the squared amplitude of the electric field
\begin{eqnarray}
\left|{\bf E}_{\rm tot}\right|^2&=&\frac{E_0^2}{2}\Big[3+2\cos ({\bf b}_1\cdot {\bf r}+\phi_{23})+2\cos ({\bf b}_2\cdot {\bf r}+\phi_{31})+2\cos ({\bf b}_3\cdot {\bf r}+\phi_{12})\nonumber\\
&&+\cos (2{\bf k}_1\cdot {\bf r}-2\omega t+2\phi_{1})+\cos (2{\bf k}_2\cdot {\bf r}-2\omega t+2\phi_{2})+\cos (2{\bf k}_3\cdot {\bf r}-2\omega t+2\phi_{3})\nonumber\\
&&+2\cos (({\bf k}_2+{\bf k}_3)\cdot {\bf r}-2\omega t+\phi_{2}+\phi_3)+2\cos (({\bf k}_3+{\bf k}_1)\cdot {\bf r}-2\omega t+\phi_{3}+\phi_1)\nonumber\\
&&+2\cos \left(({\bf k}_1+{\bf k}_2)\cdot {\bf r}-2\omega t+\phi_{1}+\phi_2\right)\Big],
\end{eqnarray}
where ${\bf b}_i=\epsilon_{ijk}({\bf k}_j-{\bf k}_k)$. Since the frequency of light is quite large, only the time-averaged value of $|{\bf E}_{\rm tot}|^2$ can affect atoms. Therefore, by dropping the terms containing $2\omega t$, we obtain a periodic dipole potential
\begin{eqnarray}
V({\bf r})=V_0\left(\frac{3}{2}+\cos ({\bf b}_1\cdot {\bf r}+\phi_{23})+\cos ({\bf b}_2\cdot {\bf r}+\phi_{31})+\cos ({\bf b}_3\cdot {\bf r}+\phi_{12})\right).
\label{Vr}
\end{eqnarray}
Red-detuned lasers give $V_0<0$ and we obtain a regular triangular-lattice potential, while the maxima of $V({\bf r})$ form a triangular lattice in the case of blue-detuned lasers ($V_0>0$)~\cite{lee-09s}. Therefore, we can cancel out unwanted sites of a triangular lattice with $V_0<0$ by overlaying another potential with $V_0>0$ so that the total potential minima form a kagome lattice~\cite{jo-12s}.

One can move the lattice potential by introducing time-dependent phase differences $\phi_{ij}(t)$ through a small frequency detuning $\delta\omega$. Let us say that we detune one of the three lasers making a triangular lattice as $\phi_1=\delta \omega t$ and $\phi_2=\phi_3=0$. In this case, we can rewrite Eq.~(\ref{Vr}) as
\begin{eqnarray}
V({\bf r})&=&V_0\left(\frac{3}{2}+\cos ({\bf b}_1\cdot {\bf r})+\cos ({\bf b}_2\cdot {\bf r}-\delta \omega t)+\cos ({\bf b}_3\cdot {\bf r}+\delta \omega t)\right)\nonumber\\
&=&V_0\left(\frac{3}{2}+\cos \left(\sqrt{3}k y\right)+2\cos \left(\frac{3}{2}k\left(x+\frac{2\delta\omega }{3k}t\right)\right)\cos \left(\frac{\sqrt{3}}{2}ky\right)\right),
\end{eqnarray}
which means that the potential moves in the $\Gamma$-$K$ direction with a constant velocity $2\delta\omega /3k$. A moving kagome optical lattice can be obtained by moving both the overlaid triangular lattices with a same velocity. Note that one uses two sets of lasers with different values of $k$ to create a kagome lattice~\cite{jo-12s}. Therefore, the frequency detuning $\delta\omega$ of each triangular-lattice potential has to be tuned so that the velocities $2\delta\omega /3k$ take a same value. 

\subsection{\label{3s}C. Effect of a moving optical lattice}

We show that the effect of a moving optical lattice in the laboratory
frame can be properly described by imposing pair formation with nonzero
center-of-mass momentum in the frame moving with the lattice potential.
In the following argument, we assume a square optical lattice for
simplicity. Its extension to multiple-sublattice geometries is straightforward.
\par
Let us start with the standard attractive Hubbard model:@
\begin{eqnarray}
\hat{H}=
-t\sum_{\langle i,j \rangle,\sigma}
\left(\hat{c}^{\dagger}_{i\sigma} \hat{c}_{j\sigma}+{\rm h.c.}\right)
-U\sum_{ i }
 \hat{n}_{i\uparrow}\hat{n}_{i\downarrow}
-\mu \sum_{i,\sigma} \hat{n}_{i\sigma}.\label{Hd}
\end{eqnarray}
In the laboratory frame $\Sigma'$, the effect of the lattice potential
moving with a constant velocity $\bf v$ can be conveniently taken into
account by the transformation that imposes a phase gradient on the
fermion operator $\hat{c}_{i\sigma}\rightarrow\hat{c}_{i\sigma}'e^{-im{\bf v}\cdot {\bf
r}_{i}}$ \cite{Levs},
and the Hamiltonian can be written as 
\begin{eqnarray}
\hat{H}'&=&
-t\sum_{\langle i,j \rangle,\sigma}
\left(e^{-i{\bf Q}\cdot {\bf r}_{ij}/2}\hat{c}^{\prime\dagger}_{i\sigma} \hat{c}'_{j\sigma}+{\rm h.c.}\right)
-U\sum_{ i }
 \hat{n}_{i\uparrow}'\hat{n}_{i\downarrow}'
-\mu \sum_{i,\sigma} \hat{n}_{i\sigma}'\nonumber\\
&=&\sum_{{\bf k},\sigma}\varepsilon_{{\bf k}+{\bf Q}/2}\hat{c}'^{\dagger}_{{\bf k}\sigma} \hat{c}'_{{\bf k}\sigma}-U\sum_{ i } \hat{n}_{i\uparrow}'\hat{n}_{i\downarrow}'-\mu \sum_{i,\sigma} \hat{n}_{i\sigma}',
\end{eqnarray}
where ${\bf Q}=-2m{\bf v}$ and $\varepsilon_{{\bf k}}=-2t(\cos k_x+\cos k_y)$. 
In the stationary state, since the condensate can coherently transport
through the lattice potential, the condensate is at rest in the
laboratory frame and consequently pairs have zero center-of-mass momentum
\begin{eqnarray}
\Delta'=U\langle\hat{c}_{i\downarrow}'\hat{c}_{i\uparrow}'\rangle=\frac{U}{M}\sum_{\rm
 k}\langle\hat{c}_{-{\bf k},\downarrow}'\hat{c}_{{\bf
 k},\uparrow}'\rangle. 
\label{Delta_lab}
\end{eqnarray}
As a result, the Hamiltonian within the Hartree-Fock-Gor'kov (HFG) mean-field approximation
takes the form 
\begin{eqnarray}
\hat{H}'_{\rm HFG}
=\sum_{\bf k}(c_{{\bf k},\uparrow}'^\dagger,c'_{-{\bf k},\downarrow})
\left(
\begin{array}{cc}
\tilde\xi_{{\bf k}+{\bf Q}/2} & -\Delta' \\
-\Delta' & -{\tilde \xi}_{-{\bf k}+{\bf Q}/2}
\end{array}
\right)
\left(
\begin{array}{c}
c'_{\bf k,\uparrow}\\
c_{-{\bf k},\downarrow}'^\dagger
\end{array}
\right),
\label{H_lab}
\end{eqnarray}
where $\tilde{\xi}_{\bf k}=\varepsilon_{{\bf k}}-\mu-Un/2$.
\par
On the other hand, In the frame $\Sigma$ moving with the optical lattice potential, the
stationary state described above is equivalent to the condensate
carrying finite flow momentum $\bf Q$ described by the order parameter 
\begin{eqnarray}
\Delta e^{i{\bf Q}\cdot{\bf r}_i}=U\langle\hat{c}_{i\downarrow}\hat{c}_{i\uparrow}\rangle=e^{i{\bf Q}\cdot{\bf r}_i}\frac{U}{M}\sum_{\rm
 k} \langle\hat{c}_{-{\bf k}+{\bf Q}/2,\downarrow}\hat{c}_{{\bf k}+{\bf
 Q}/2,\uparrow}\rangle.
\label{Delta_lat}
\end{eqnarray}
Within the HFG mean-field approximation, the Hamiltonian~(\ref{Hd})
takes the form
\begin{eqnarray}
\hat{H}_{\rm HFG}
=\sum_{\bf k}(c_{{\bf k}+{\bf Q}/2,\uparrow}^\dagger,c_{-{\bf k}+{\bf
Q}/2,\downarrow})
\left(
\begin{array}{cc}
{\tilde \xi}_{{\bf k}+{\bf Q}/2} & -\Delta \\
-\Delta & -{\tilde \xi}_{-\bf k+\bf Q/2}
\end{array}
\right)
\left(
\begin{array}{c}
c_{{\bf k}+{\bf Q}/2,\uparrow}\\
c_{-{\bf k}+{\bf Q}/2,\downarrow}^\dagger
\end{array}
\right),
\label{H_lat} 
\end{eqnarray}
In the paper, we set up flow in the frame $\Sigma$ imposing pair formation
with nonzero center-of-mass momentum as in Eq.~(\ref{Delta_lat}).
\par
The mean-field Hamiltonians in Eqs.~(\ref{H_lab}) and (\ref{H_lat}) as
well as the order parameters in Eqs.~(\ref{Delta_lab}) and (\ref{Delta_lat}) are
equivalent under the shift of the origin of momentum of the operators: ${\hat c}'_{\bf
k,\sigma}\leftrightarrow\hat{c}_{{\bf k}+{\bf Q}/2,\sigma}$.
This shows that the effect of the moving optical lattice in the frame
$\Sigma'$ can be properly described by imposing pair formation with
nonzero center-of-mass momentum $\bf Q$ in the frame
$\Sigma$.

\end{document}